\documentclass[12pt]{article}
\usepackage{amsmath,amssymb,graphicx,mathrsfs,hyperref}

\newcommand{\be}{\begin{equation}}
\newcommand{\ee}{\end{equation}}
\newcommand{\bea}{\begin{eqnarray}}
\newcommand{\eea}{\end{eqnarray}}

\def\({\left(} \def\){\right)}

\begin{document}
\title{\vspace{-1.8in} \begin{flushright} {\footnotesize CERN-PH-TH/2011-204}  \end{flushright}
\vspace{3mm}
\vspace{0.3cm} {\Large Non-perturbative unitarity constraints \\ on the ratio of shear viscosity to entropy density \\ in UV complete theories with a gravity dual}}
\author{\large Ram Brustein${}^{(1,2)}$,  A.J.M. Medved${}^{(3,4)}$ \\
 \hspace{-1.5in} \vbox{
 \begin{flushleft}
  $^{\textrm{\normalsize
(1)\ Department of Physics, Ben-Gurion University,
    Beer-Sheva 84105, Israel}}$  \\ $^{\textrm{\normalsize
 (2)  CERN, PH-TH Division, CH-1211, Gen\`eve 23,  Switzerland}}$  \\
$^{\textrm{\normalsize (3)  Department of Physics \& Electronics, Rhodes University,
  Grahamstown 6140, South Africa }}$ \\
$^{\textrm{\normalsize (4)  School of Physics, KIAS,
  Dongdaemun-gu, Seoul 130-722,  Korea}}$ \\ \small \hspace{1.7in}
    ramyb@bgu.ac.il,\  j.medved@ru.ac.za
\end{flushleft}
}}
\date{}
\maketitle

----------------------------------------------------------------------

\begin{abstract}
We reconsider, from a novel perspective, how unitarity constrains the corrections to the ratio of shear viscosity $\eta$ to entropy density $s$. We start with higher-derivative extensions of Einstein gravity in asymptotically anti-de Sitter spacetimes. It is assumed that these theories are derived from string theory and thus have a unitary UV completion that is dual to a unitary, UV-complete boundary gauge theory. We then propose that the gravitational perturbations about a solution of the UV complete theory are described by an effective theory whose linearized equations of motion have at most  two time derivatives. Our proposal leads to a concrete prescription for the calculation of $\eta/s$ for theories of gravity with arbitrary higher-derivative corrections. The resulting ratio can take on values above or below $1/4\pi$ and is consistent with all the previous calculations, even though our reasoning is substantially different. For the purpose of calculating $\eta/s$, our proposal also leads to only two possible candidates for the effective two-derivative theory: Einstein and  Gauss-Bonnet gravity. The distinction between the two is that Einstein gravity satisfies the equivalence principle, and so its graviton correlation functions are polarization independent, whereas the Gauss-Bonnet theory has polarization-dependent correlation functions.  We discuss the graviton three-point functions in this context and explain how these can provide additional information on the value of $\eta/s$.
\end{abstract}
\newpage

\section{Introduction}

The gauge--gravity duality \cite{maldacena1,Maldacena}
can  be used to learn about the hydrodynamic properties of strongly coupled gauge theories in the limit that their number of colors $N$ is infinite.  This is achieved through the study of  weakly coupled anti-de Sitter (AdS) gravity theories of one dimension higher  \cite{KSS-hep-th/0309213,SS}.

In the case that the theory in the AdS bulk is Einstein gravity,
the ratio of the shear viscosity ($\eta$) to the entropy density ($s$)
in the dual gauge theory takes on  the well-known  value of
\cite{PSS,PSS2,KSS-hep-th/0309213,newreview}
\be
\frac{\eta}{s}\;=\; \frac{1}{4\pi} \;
\label{kss}
\ee
in natural units. Kovtun, Son and Starinets
\cite{KSS} conjectured that $\frac{\eta}{s}\;\geq\; \frac{1}{4\pi}$ holds for
all strongly coupled fluids, and this restriction has since been referred to
as the KSS bound.

We are considering generic extensions of Einstein
gravity in an (asymptotically) AdS spacetime of  dimensionality
$D\geq 5$. Our focus is on static black brane backgrounds,
relevant for studying the hydrodynamics
of  the field theory (FT)  dual. We assume that there are
higher-derivative terms in the extended theory, which amounts
to finite $N$ or quantum corrections to the FT.
The number of derivatives can be arbitrary, but the extensions
are to be regarded as perturbatively small relative
to the Einstein term. Our focus is on corrections to the ratio
of two-point functions $\eta/s$; however, as a consequence of our results, we find that quantities depending on the  higher-point functions can similarly be studied.

The entropy density $s$ can be directly obtained from the
Wald formula \cite{wald1,wald2}, whereas the viscosity $\eta$ can be determined (at least in principle) by a suitable generalization of the Kubo formalism. The calculation of the ratio $\eta/s$ for gravity theories having two-derivative linearized equations of motion is well understood. We refer to such theories as
``two-derivative gravity'' or sometimes as the ``Lovelock class''. These are the theories that belong to the special class of Lovelock theories \cite{LL} or are,  like $f({\cal R})$ gravity, closely  related.

But, if the linearized equations of motion contain terms  with four or more derivatives, the theory inevitably contains Planck-scale ghosts and it is not clear whether these ghosts affect the value of the hydrodynamic parameters. So  the situation is more complicated than often appreciated, and this matter has not yet  been fully resolved. This is the aim  of the current paper.

This issue of ghosts  has been discussed in \cite{BMunit},
where   $\eta/s$ was calculated for the FT duals of four-derivative gravity theories. The ghost modes and their influence on the computed values of $\eta/s$ were explicitly exposed and understood. Then a set
of boundary conditions (BC's) that eliminates the ghost modes was proposed. It was further claimed that, once these BC's are enforced, the  validity of the KSS bound is ensured.

The conclusion of \cite{BMunit} can be confronted with the particular case of Gauss--Bonnet gravity. This is a special theory for which the Lagrangian contains four-derivative terms but the linearized equations of motion contain at most two derivatives,  and so  it is two-derivative gravity by our definition. For this case, it is well known that the ratio $\eta/s$ can be below the KSS bound. Apparently, by having no ghosts to begin with, the Gauss--Bonnet model is able  to evade the need for enforcing the BC's of \cite{BMunit} while  violating the bound.

Here, we extend the considerations of \cite{BMunit} and propose
that  metric perturbations about a solution of string theory are {\it always} described by two-derivative gravity.
This should apply,  in particular, to black branes in AdS space. So that, given a higher-derivative theory, the resulting effective two-derivative theory should  be used for {\it all} calculations of the two-, three-  (and  higher-) point functions.
This includes (but is not limited to)  $s$ and $\eta$, as these parameters are directly  related to the two-point functions of
specific gravitons in the black brane background.

This restriction to two-derivative gravity, for which there
is a  limited number of choices, implies that all such gravity theories  and their FT duals depend on only a small number of parameters.
Moreover, the above proposal already provides a clear distinction in
philosophy between what we are suggesting and the previous methods for calculating $\eta/s$. The resulting  two-derivative theory should not be
viewed as a perturbative extension of Einstein gravity (even if
its higher-derivative origin is) since one uses the information from the UV-complete parent theory to allow the propagator to change.  And so the arguments for enforcing the KSS bound in \cite{BMunit} no longer apply.

The rest of the paper proceeds as follow.  Section~2 starts  with
an explanation on how to  extract the viscosity from the one-particle irreducible  (1PI) effective action, with special attention
to the implications of higher-order corrections. We then  propose a novel set of BC's that maintains explicit unitarity by reducing any higher-derivative theory to an effective two-derivative model. It is made clear how our proposal  differs from previous methods, which are either explicitly or implicitly based upon  a Wilsonian form of the effective action. Next, the entropy density is considered; in particular, the application of the standard calculations of $s$ to a higher-derivative theory. We are, at the end, able to prescribe the ratio $\eta/s$ for a higher-derivative theory in an unambiguous way.

In Section~3, we explain that, for current purposes, any such  two-derivative model must be either an Einstein theory or a Gauss--Bonnet theory. The value of $\eta/s$ is shown to be  sensitive to this choice and we provide a criterion to determine which theory should be used. This criterion relies on the distinction between the Einstein and Gauss--Bonnet forms of the two-, three- (and higher-)
point functions.

The final  section  briefly summarizes our conclusions
and explains their relevance to multi-particle correlations in strongly coupled fluids.

Appendix~A contains the details of a calculation presented in Section~3,
while Appendices~B and~C review  relevant known results.

\section{A prescription for evaluating $\eta/s$}

We are considering a theory of gravity with an Einstein--Hilbert term and arbitrary higher-derivative corrections. The  main objectives of this section are to explain how the 1PI effective action can be used to extract the viscosity, propose BC's that ensure  unitarity  and provide a precise prescription for calculating  $\eta/s$. An important conclusion of this section is that {\it any} such gravitational theory should be a two-derivative gravity. Although our current focus is primarily on the viscosity (and, then, $\eta/s$), a similar procedure can be applied to any graviton $n$-point function.

\subsection{Initial considerations}

We assume here that the theory describing  the AdS bulk spacetime  is
string theory,  a unitary and UV-complete theory.  In the low-energy limit, as relevant for calculating transport coefficients like $\eta$, string theory can be approximated by an effective gravitational action, which includes the lowest-order (Einstein) term and possibly some higher-derivative corrections.

We further assume that  the full UV-complete theory has a  black brane
solution. We do not need to know the exact form of this solution,
only its geometry.

As a concrete example, we have in mind the case of
a  $D$-dimensional AdS  black brane for which the geometry is described by the following background metric
\be
ds^2\;=\;-F(r)dt^2+ \frac{dr^2}{F(r)}+\frac{r^2}{L^2} dx_i^2\;.
\label{geometric}
\ee
The index denotes the transverse space dimensions $i=1,\dots,D-2$, $L$ is the AdS radius of curvature and $r$ is the radial coordinate
(orthogonal to the brane).
The function $F$ vanishes on the horizon at $r=r_h$, $\;F(r_h)=0\;$,
and asymptotes  to $r^2/L^2$ as the AdS boundary
($r\to\infty$)  is approached.

According to the gauge--gravity duality,
the bulk has a UV-complete and unitary FT dual that ``lives'' at
the AdS boundary; for example, ${\cal N}=4$ supersymmetric  Yang-Mills theory \cite{maldacena1} or the Kats--Petrov class of ${\cal N}=2$ conformal (C)FT's \cite{same-day}.

We wish to study small metric perturbations about the solution of the full theory and describe them in terms of an effective field theory. Since our interest
is in physical observables in the bulk, we should calculate on-shell graviton scattering amplitudes near the boundary of the AdS spacetime. These
amplitudes  are then dual to gauge-invariant correlation functions of the
energy-momentum tensor of the FT, whereas other types of bulk amplitudes
are generally not physical.

Let us make a cautionary comment about the on-shell condition in AdS space. Since the physical quantities are really on-shell boundary amplitudes, it should have been possible
to describe them purely in terms of boundary quantities. However, in practice (as discussed in detail below), one needs to define the solutions of the linearized equations also in terms of bulk quantities. For instance, in a black brane background, one of the boundary conditions is usually imposed on the horizon. This involves gauge choices in the bulk, and one should make sure that the final results are  physical and not affected by any of these choices.

The relevant effective bulk action for the gravitons is, therefore,
the 1PI action as derived from string theory, since
this serves to   generate   the physical on-shell amplitudes.
This 1PI action
 comes already equipped with an
on-shell condition that defines the solution of the
linearized equation for gravitons around a black brane background.

\subsection{Extracting $\eta$ from the 1PI action}

Let us recall how the 1PI action is derived from string theory, starting
with the case of  ${\cal N}=1$ supergravity  in 10-dimensional
Minkowski spacetime. The on-shell condition for such theories is
\begin{equation}
\Box_E h_{\mu\nu}=0\;,
\label{onshell}
\end{equation}
up to quite high order (as explained shortly). $\Box_E$ is meant as the differential operator that satisfies the equations of motion from an Einstein theory. Corrections to the 1PI action start with terms that are quartic in the Riemann tensor (specifically, quartic contractions of
the  Weyl tensor \cite{Gross-Witten,Tsey})  and so contain 8-derivative terms. These terms determine only the 1PI four-point function (and possibly some higher-point functions) while not changing the one-,  two- and three-point functions. This is the standard outcome when evaluating the 1PI action order by order in perturbation theory; new interactions do not change the lower-point functions. In particular, they do not change the one-point function (equivalently, the on-shell condition)  nor the two-point function (equivalently, the propagator).

When one considers the 1PI effective action in a lower-dimensional AdS
Schwarzschild geometry ({\it e.g.}, $D=5$), the situation changes. Because
the spacetime is no longer flat, the Weyl tensor is no longer vanishing in the new geometry. So that, if one computes the corrections coming from the quartic (Weyl$^4$) terms, the on-shell condition, propagator and three-point function are all modified. That the on-shell condition and the propagator are changed by interaction-induced corrections is a consequence of treating the full theory in a non-perturbative way rather than perturbatively near a specific vacuum solution. For instance, terms that only contribute to interactions in a given vacuum can modify the propagator in another. This is the main difference between the perturbative treatment discussed in \cite{BMunit} and the current non-perturbative treatment.

The FT viscosity $\eta$ is a physical quantity and is, therefore,  defined by a physical on-shell amplitude of a specific class of perturbations,
the transverse--traceless gravitons $h_{xy}$.~\footnote{The labels $x$ and $y$  denote transverse brane directions that are
mutually orthogonal as well as orthogonal to
the direction of propagation.} Of course, the viscosity $\eta$ can be defined, in principle, purely in terms of a FT  prescription that is completely independent of the bulk.

Let us now review how the on-shell condition is enforced in a black brane
geometry for the case of the bulk theory being Einstein gravity. Then we will discuss the on-shell condition for more complicated situations.

First, the Einstein equation for the gravitons is linearized in some gauge. The conventional choice is the radial gauge, which sets all graviton perturbations
$h_{\mu r}$ to zero such that $\mu$ is arbitrary.~\footnote{In the radial gauge,  the $\{x,y\}$ gravitons decouple from the other polarizations and, so, this gauge is compatible with transverse and traceless $h_{xy}$'s.}
The resulting linearized form of the Einstein equation is
$\;\Box_E h_{\mu\nu} =0\;$.

The next step is  to choose  the
BC's. The standard choice is Dirichlet BC's on the boundary of AdS  and
incoming BC's at the horizon of the black brane. These choices  completely
fix the solution of the linearized equations and, thereby, uniquely determine
the on-shell conditions for gravitons in this background.

The solutions of the linearized equations are of the plane-wave form
$\;h_{\mu\nu} \sim h_{\mu\nu}(r) e^{i \vec{P}\cdot{\vec{X}} - i \Omega T}\;$,
 where  $\vec{X}$ are the space coordinates and $T$ is the time coordinate for
both the FT and the brane. $\vec{P}$ is the dimensionless space momentum and
$\Omega$ is the dimensionless frequency; both of which are
in units of the brane temperature.

One then  uses the solution of the linearized equation to evaluate the
on-shell value of the two-point function of gravitons near the AdS boundary.
By the  standard rules of the AdS/CFT duality, this correlator is  related
to   the gauge-invariant two-point function of the
energy-momentum tensor of the FT.
Since the FT (as well as  the bulk) is in a thermal state, this on-shell
two-point function is also the 1PI retarded propagator.

The viscosity $\eta$ and the other transport coefficients of the FT can be extracted from the two-point function of the energy-momentum tensor or, equivalently, from that of the  gravitons near the AdS boundary in  the hydrodynamic limit. In this limit,  the frequency $\Omega$ and the momentum $P$ both vanish,
possibly at different rates.

In practice, the value of $\eta$ can  be evaluated  by
a number of means. For instance,  via the Kubo formula \cite{PSS},
the membrane paradigm \cite{KSS-hep-th/0309213} or
simply by looking at the two-point function $\langle h_{xy}h_{xy} \rangle$  at the horizon (equivalently,  at the associated kinetic terms in either the action or the equations of motion)
\cite{BMratio,IqLu,CAI}. The latter  is so because, in the hydrodynamic limit and when the equations of motion contain at most two derivatives, the effective action has the same value at the horizon  and  at the boundary.

\subsection{Extracting $\eta$ from higher-derivative gravity}

Let us now discuss how the on-shell condition is enforced when the lowest-order
Einstein--Hilbert action is corrected by higher-order terms (such as the previously discussed quartic terms).  To fix the on-shell condition here, we will use exactly the same principles that were used above in the simpler case. A slight complication arises  when the on-shell condition gets corrected at higher orders. In which  case, there is some additional freedom in defining the on-shell condition due to field redefinitions adding terms proportional to the lowest-order equations. This freedom will play a significant role in  what follows.

To see that the on-shell condition for the gravitons is modified by the presence of higher-order corrections,
we can look at the linearized equation for the $h_{xy}$ gravitons.
As discussed in \cite{wald2} and clarified in \cite{BGHM},
the Lagrangian of a general extension of Einstein gravity can be expressed  in terms of the metric, the Riemann tensor and its symmetrized covariant
derivatives,  while anti-symmetric combinations
of derivatives can  be traded off for more Riemann tensors,
$\;{\cal L}={\cal L}\left[g_{ab},{\cal R}_{abcd},
\nabla_{a_1}{\cal R}_{abcd},
\nabla_{\left(a_1\right. }\nabla_{\left. a_2\right)}{\cal R}_{abcd},\ldots\right]\;$.
The linearized equation for the $h_{xy}$ gravitons can be
obtained by  expanding the full equations of motion
\be
\frac{\partial{\cal L}}{\partial g^{pq}}\;-\;
2\nabla_a\nabla_b{\cal X}^{a\;\;\;\;b}_{\;\;pq\;\;}
\;+\;{\cal R}_{abcp}{\cal X}^{abc}_{\;\;\;\;\;q}
\;-\; \frac{1}{2} g_{pq}{\cal L}
\;=\; 0\;
\label{wald-EQ}
\ee
to linear order in $h_{xy}$. Here,
\bea
{\cal X}^{abcd}\;\equiv\; \frac{\partial{\cal L}}{\partial {\cal R}_{abcd}}
 \;&-&\;\nabla_{a_1}\left(\frac{\partial{\cal L}}
{\partial[\nabla_{a_1} {\cal R}_{abcd}]}\right) \nonumber \\
\;&+&\;\nabla_{\left(a_1\right. }\nabla_{\left. a_2\right)}
\left(\frac{\partial{\cal L}}
{\partial[\nabla_{\left(a_1\right.} \nabla_{\left. a_2\right)}{\cal R}_{abcd}]}\right)
\;+\; \ldots\;
\label{W2}
\eea
and the ellipsis means  ever-increasing numbers of
symmetrized derivatives.

The linearized equation can also be regarded as a perturbative expansion
in derivatives. Schematically, suppressing tensorial structures and indices,
\be
\Biggl(\Box_E  + \epsilon c_2(r) \nabla\nabla  + \epsilon L c_3(r)
(\nabla)^3
+ \epsilon L^2 c_4(r)(\nabla)^4
\;+\ldots\;
\nonumber
\ee
\be
\dots\; +\epsilon^{n-1} c_{2n-1}(r) L^{2n-3}(\nabla)^{2n-1}
+\epsilon^{n-1} c_{2n}(r) L^{2n-2}(\nabla)^{2n} +\;\ldots \Biggr)h_{xy}
  \;=\; 0\;,
\label{genexp}
\ee
where $\Box_E$ is the operator for the Einstein equation
($\;\Box_E h_{ab}=0\;$), $\epsilon\ll 1\;$ is a dimensionless
perturbative parameter controlling
the strength of the corrections and
the dots  denote increasing numbers of derivatives
 as well as mass terms of pertrubative order.~\footnote{One can
absorb such mass terms into the definition of $\Box_E$,
so that these are essentially irrelevant. Further, one can disregard
perturbative terms with  a single derivative acting on a graviton, as these
are related to mass terms by integration by parts.}
The tensors,  $c_2(r)$, $c_3(r)$, {\it etc.},  are a model-dependent collection of dimensionless radial tensors that are at most of order unity but can have subleading contributions of order $\epsilon$ (and higher)  from yet higher-derivative corrections. These tensors represent the most general way to contract a given number of covariant derivatives and are built with the background metric, background Riemann tensor and its symmetrized derivatives.
Consistency of the perturbative expansion fixes the lowest power of $\epsilon$ for a given number of derivatives.
For classical $\alpha'$ string corrections,
which correspond to finite $N$ corrections for the FT,
$\;\epsilon \sim l_p^2/L^2$ (with $l_p$ being the Planck length).
On the horizon, $\Box_E$ reduces to $\Box$, the standard d'Alambertian.

Let us recall from \cite{BMunit} how this procedure works in the simplest case
of four-derivative corrections to Einstein gravity.
 Eq.~(\ref{genexp})
then has the simpler form~\footnote{This is only strictly true in
the hydrodynamic limit. See \cite{BMunit} for clarification.}
 (up to henceforth implied  mass terms)
\be
\Box_E h_{xy} + \epsilon b(r) \Box_E h_{xy} + \epsilon L^2 a(r) \Box^2_E h_{xy}
 \;=\; 0\;.
\label{gen1}
\ee

To summarize the preceding discussion, the linearized equation for the
$\{x,\ y\}$ gravitons is generically of the form
\begin{equation}
(\Box_E+ \epsilon {\mathcal D} + \dots) h_{xy} =0\;,
\label{highonshell}
\end{equation}
where ${\mathcal D}$ denotes the leading-order higher-derivative corrections
to the linearized equation
(coming from the leading corrections to the Einstein--Hilbert action). The exact form of ${\mathcal D}$ will not be essential for what follows.

The operator ${\mathcal D}$ generically contains  terms with  four
or more derivatives, including some  with at least four time derivatives.  The equation will include more and more time derivatives as the perturbative order increases.
So that, to define the on-shell condition,  Eq.~(\ref{highonshell}) has to be supplemented with new BC's in addition  to  those already specified in the Einstein case. Recall that some of the BC's are fixed at the boundary of AdS
and others on the horizon of the black brane.
The most relevant BC's for the current discussion are those at the outer
boundary.

That Eq.~(\ref{highonshell}) contains more than two time derivatives is  very problematic because of the ghost modes that must appear in such cases. Since our treatment is classical, the problem manifests itself in severe instabilities if the ghost modes have non-vanishing amplitudes. (See \cite{BMunit} for additional discussion on this point.) What makes the situation a little  more bearable is that the ghosts are typically at the cutoff (Planck) scale and,
therefore, expected to be irrelevant to the calculation of hydrodynamic  coefficients like $\eta$.

This situation when ghosts are present but  have a cutoff-scale mass is ambiguous in connection to the calculation of $\eta$ (and any other coupling).  One might  reason that it is  safe to disregard their presence.  However, to proceed, one should in any case ensure that the ghosts
have decoupled before the computation is carried out. Rather than rely on a general argument that may or may not apply, we propose to exorcize the ghosts from the get go; ensuring that they do not have the opportunity to interfere with the low-energy physics.

In fact, by the same line of reasoning, it is safer to eradicate all the extra modes with cutoff-scale masses, ghosts or otherwise. We propose to accomplish this goal through a particular choice of BC's. The rational for this choice is  that, if some properties of the cutoff-scale modes determine a low-energy physical quantity, then the resulting value of this quantity cannot be trusted. So that, whenever a calculation involves the cutoff-scale modes,  one should make sure that their amplitude, couplings, {\it etc.}, do not enter into the final result. The best way to achieve this is, in our opinion,  to set them to zero from the very start.

We are now ready to make our central assumption about how the BC's should be chosen. We propose that

\begin{equation}
\parbox{4in}{\it
The boundary conditions imposed on the higher-derivative equation have to be chosen in a unique way such that, from the many solutions to the higher-order equation (\ref{highonshell}), only one  mode has a non-vanishing amplitude. This mode is the unique mode that, as $\epsilon$ vanishes, connects  continuously
to the solution of  the lowest-order equation (\ref{onshell}) when
supplemented with its original boundary conditions.}
\label{mainassumption}
\end{equation}

\ \\

The essential difference between the proposal (\ref{mainassumption}) and the proposal in \cite{BMunit} is that, here,   we are proposing to set to zero the amplitude of {\it all} the cutoff scale modes and not just the ghosts.
For four-derivative theories, the two proposals are equivalent.

Let us now describe how the proposal (\ref{mainassumption}) is implemented  for gravitons in an AdS black brane background in the presence of corrections to Einstein gravity. To implement our choice of BC's on Eq.~(\ref{highonshell}), let us suppose (for simplicity) that, at leading order in $\epsilon$,
the operator ${\mathcal D}$
is obtained from a four-derivative
or Riemann$^2$ correction. In \cite{BMunit}, it was shown that it
is  possible to factor Eq.~(\ref{highonshell}) in the following way:
\begin{equation}
\Box_E+ \epsilon {\mathcal D} =  \left[1+ \epsilon b_2(r)+ \epsilon a(r) \Box_E\right]
\left[1+ \epsilon b_1(r)\right] \Box_E\;,
\label{factor}
\end{equation}
where $a(r)$ and  $b(r)=b_1(r)+b_2(r)$
are functions of the AdS radial coordinate
that are  uniquely determined by the operator ${\mathcal D}$.
To verify this form, one need only expand out the right-hand side
to first order in $\epsilon$ and compare it  with
Eq.~(\ref{gen1}).~\footnote{There are subtleties regarding the commutation
of derivatives; see \cite{BMunit}. Alternatively,
one can bypass this issue by imposing the limit of large (radial) momenta
on the gravitons, so that the coefficients are effectively constant.}

From Eq.~(\ref{factor}), we can extract the two modes
 that the corrected theory contains (now
suppressing  graviton indices): one massless graviton $h_1$ and one  additional mode $h_2$ whose mass is near the cutoff scale $1/\epsilon$. Only one additional mode is added in this case because the highest derivative is of fourth order. The distinction between the modes is still subject to the aforementioned freedom of field redefinition.
Here, the freedom is  manifested in the functions  $b_1$ and $b_2$,
which are arbitrary provided that their sum
remains
fixed. This freedom cannot affect physical quantities and indeed it does not.

For a fixed identification of modes, the BC's  are imposed according to
(\ref{mainassumption}) as follows: The BC's for the massive  solution $h_2$,
\begin{equation}
[1+ \epsilon b_2(r)+ \epsilon a(r) \Box_E]h_2 =0
\end{equation}
are $h_2 =0$ at the boundary and incoming at the horizon. This constrains
the amplitude of $h_2$ to vanish everywhere in the bulk (and obviously at the AdS boundary)  by setting both its normalizable and non-normalizable modes to zero at the boundary. For the massless graviton, one chooses the ``standard BC's",
incoming at the brane horizon and Dirichlet at the AdS boundary.

The 1PI effective action to a given order in gravitons is obtained by expanding the original higher-derivative effective action to the same  order and,
then, setting the extra mode to zero as required from the solution of its equations of motion. The result is  a two-derivative action for the (modified) massless graviton. It is still not uniquely defined because of the freedom to change $b_1$, while keeping  $b_1+b_2$ fixed.  Ultimately, though,  physical quantities must
 not depend on this freedom (as argued above).

The situation in which the equation contains terms with higher than four derivatives is conceptually the same as we have discussed, although technically more complicated. The factorization procedure is expected to follow along similar lines and produce an additional factor for any additional pair of derivatives. Out of all the modes, one still has to choose the single mode that connects continuously to the massless graviton and set BC's such that all the rest vanish.
The theory must  then necessarily reduce to quadratic order or
\be
\left(1 + \epsilon b(r)\right) \Box_E h_1  \;=\; 0\;,
\ee
where the  $b(r)$ here is completely analogous (and equally ambiguous) to
the  coupling in the four-derivative case.
 We have not, however, implemented the proposal in a concrete example derived from a theory with six or more derivatives.

Once the excess modes have been eradicated (and the freedom in $b$ has been resolved), what is left
is a two-derivative theory for which the usual AdS/CFT rules  apply. One determines any $n$-point function
by turning on a source (the non-normalizable mode) for the``normal" graviton,
taking the $n$'th derivative of the action with respect to this source and
then evaluating   on the AdS boundary (with some prescription for removing divergences). Since $\eta$ is expressed in terms of a 1PI two-point function, which is also the two-point function of the
FT energy-momentum tensor,
the proposed procedure then computes $\eta$  in an unambiguous way.

The very same reasoning extends
to all  other transport coefficients and higher-point functions.

The main conclusion of this section is that, when the proposed BC's are
enforced, the resulting 1PI effective action for the remaining mode
is a two-derivative gravity (meaning that the linearized equations of motion derived from it contain at most two derivatives).

\subsection{Extracting $\eta$ from the Wilsonian effective action}

Another way of solving Eq.~(\ref{highonshell}) and calculating $\eta$ is via the Wilsonian effective action.  In practice, the calculations are only presented at the level of the equations of motion and done in the ``reduced-action method" of Banerjee and Dutta \cite{BD}.
All other accepted methods  ({\it e.g.}, \cite{ALEX}--\cite{CLEV})
are equivalent to it.

By this method, one iteratively reduces the higher-derivative equation to a  two-derivative form as follows: The lowest-order field equation (\ref{onshell})  can be written explicitly as
(suppressing graviton indices)
\begin{equation}
\left[\partial^2_r  + f_1(r) \partial_r + f_2(r)\right] h(r)=0 \;.
\label{BD1}
\end{equation}
The functions $f_1$ and $f_2$ depend on the background and can also depend on $P$, $\Omega$ and $T$. Their explicit form is not relevant for this discussion.

Eq.~(\ref{BD1}) is then used iteratively to reduce the order of the operator ${\mathcal D}$ in Eq.~(\ref{highonshell}) until the full equation reduces to a quadratic equation. For example, when the operator ${\mathcal D}$ has
a four-derivative term as its higher order, then one has to use Eq.~(\ref{BD1}) twice. Clearly, the procedure amounts to integrating out the massive modes, and so the equations that the massless graviton obeys can be derived from the corresponding Wilsonian effective action.

In our notation, each iterative step is equivalent to
using the relation
\be
\Box_E h \;=\; {\cal O}(\epsilon) h\;.
\ee
Then, for example,
\be
\Box^2_E h  \;=\; \Box_E \left[\Box_E h \right]
\;\to\; \Box_E \left[{\cal O}(\epsilon) h \right]
\;\sim\; \epsilon \Box_E h \;.
\ee
One does this until there are at most two-derivative terms left and then discards any subleading term in $\epsilon$.
The  final result of this iterative process is then to reduce
the higher-derivative Eq.~(\ref{highonshell}) to a quadratic
in derivatives.

For instance, when applied to the field equation (\ref{gen1})
of a four-derivative corrected theory, this reduction process
gives back simply
\cite{BMunit}
\be
\left(1 + \epsilon b(r)\right) \Box_E h  \;=\; 0\;.
\label{BDXXX}
\ee
Eq.~(\ref{BDXXX}) is then solved, as usual, with Dirichlet BC's
at the AdS boundary and  incoming BC's at the black brane horizon.
Eq.~(\ref{BDXXX}) can also be utilized to find a quadratic (Wilsonian)
effective action, which  may then be used
(according to the standard AdS/CFT rules) to calculate bulk amplitudes
near the boundary of AdS or, equivalently,
 FT correlation functions.

In short,  one ends up with \cite{BMunit}
\be
\eta \;\propto\; 1+\epsilon b(r_h)\;,
 \ee
which can be verified by extrapolating
the usual two-derivative methods.
(The constant of proportionality is fixed by Einstein's theory.)
But, importantly,  $b(r)$ is the exact same
$b(r)=b_1(r) + b_2(r)$ that
appeared in the non-unique decomposition in Eq.~(\ref{factor}).

\subsection{Comparison of the 1PI and Wilsonian approaches}

The Wilsonian calculation can be compared explicitly to the 1PI calculation with our proposed BC (\ref{mainassumption}). The result  is that the Wilsonian approach leads to a solution with a non-vanishing, order-$\epsilon$ amplitude for the redundant mode.

The two procedures  give the same equation for the gravitons up to order $\epsilon$,  and only differ at order $\epsilon^2$. Meaning that, if one wishes to calculate to order $\epsilon$, it is possible to use either method. Since, in general,
the Wilsonian effective action is not well defined in string theory, the justification for its use comes from the fact that it is expected to (and indeed does) lead to results that are equal to the ones obtained from the 1PI effective action at each order in perturbation theory. However, at higher orders this may become a delicate balancing act.

The final result is that both methods agree on the value of $\eta$ (ambiguity aside) to leading order in $\epsilon$.  This is somewhat of a red herring, as both $\eta$ and $s$ are densities and, as such, depend on the choice of coordinates, field definitions and so forth. Only the ratio $\eta/s$ is an invariant physical quantity. Meaning that it is now time to bring $s$ into the discussion,
which we do next.

\subsection{Calculating the entropy density}

The entropy density $s$ is, similarly to $\eta$, related to a graviton two-point function \cite{BGH-0712.3206,BGHM}, and so  its calculation
 is subject to some of the same  considerations.

Let us momentarily consider  a two-derivative
theory  of gravity that is not necessarily Einstein's theory.
It is by now established \cite{BMratio,IqLu,CAI} that the extraction of $\eta$ amounts to  reading off the horizon value of  $b$ in an equation formally identical to (\ref{BDXXX}).
Perhaps less well known is that one can play the exact same game
to extract the entropy $S$ or its density $s$. In this case, the gravitons
in question are the $h_{rt}$ gravitons, which decouple from all
others at the horizon.\footnote{The zero-modes of the $h_{rt}$ gravitons do not vanish on the horizon despite the choice of radial gauge. This gauge is inconsistent with the identity $\;h_{ab}=\nabla_a\chi_b+ \nabla_b\chi_a\;$
 ($\chi^a$ is the Killing vector),  which
has to hold on the horizon  \cite{wald1,wald2}.
This is similar to the zero-mode problem in the Coulomb Gauge  in electromagnetism and is resolved in a similar way.\label{killing}}
The precise equivalence of this
procedure to the Wald
formulation
was first pointed out in \cite{BGH-0712.3206} and then made
explicit in \cite{BGHM}.  That is, one can look at the $h_{rt}$
component of the field equation
\begin{equation}
(1+ \epsilon b_s(r)) \Box_E h_{rt} \;=\; 0\;,
\label{BDS}
\end{equation}
and read off the horizon value $b_s(r_h)$.
This represents the correction to the
entropy density  when the Einstein value has been normalized to unity.

A subtle  difference in the $\eta$ and $s$ calculations becomes
apparent only when higher-derivative corrections are introduced.
Recall that $\eta$ has an arbitrariness that is introduced  by  the presence of  a ghost mode. On the other hand,  $s$ is somehow ``smarter''. As clarified in \cite{BGHM}, any term with four or more derivatives can  be  reduced to two derivatives (or less) via successive applications of  the Killing identity,  $\nabla_c \nabla_a \chi_b \;=\; -{\cal R}_{abcd}\chi^d \;$. This reduction process follows
from a relation between the $h_{rt}$ gravitons and the horizon Killing vector
$\chi^{a}$ that is valid only at the horizon (see footnote~{\ref{killing}).

It follows that, at the horizon $r=r_h$, the  $\{r,t\}$  counterpart of Eq.~({\ref{genexp}) for a higher-derivative theory is
\begin{equation}
(1+ \epsilon c_2(r_h))\Box h_{rt}+\dots \; =\; 0\;,
\end{equation}
where the additional terms that are represented by the ellipsis are {\it only}
$\epsilon$-order mass terms. Higher-derivative
terms do  not exist! This simplification
is unique to this class of  perturbations and makes the entropy  density special amongst graviton  couplings.

Strictly speaking, as we have just argued,  the Wald entropy density $s$ is insensitive to higher-order corrections and to the ghost modes that they induce. Thus, $s$  appears to be free of the type of ambiguities
that haunt $\eta$ and other couplings. However, in the general context of the discussion, this conclusion is misleading because it holds only at the horizon.~\footnote{Although the Wald density can only be defined at the horizon,
one can use FT thermodynamics and AdS/CFT rules to define $s$
in terms of the $h^{t}_{\ t}$ and $h_{\ \ x_{i}}^{x_{i}}$ gravitons.
This definition has been shown to limit to the Wald formula
as $r\to r_h$ \cite{BwithG}.}   As we move away from the horizon, the ghost modes and their associated instabilities come back.

Our conclusion is that, to define  $\eta$ or $s$ or any other physical coupling, the only safe way to proceed is  to restrict the discussion
to two-derivative theories.

\subsection{Calculating the ratio $\eta/s$ \label{sss}}

We have tentatively identified the viscosity for an effective two-derivative
theory as $\eta\propto 1+\epsilon b(r_h)$.
However, only a ratio of densities, such as $\eta/s$ is  physical.
The ratio $\eta/s$ is gauge invariant and also invariant under bulk field redefinitions. The most convenient way to evaluate it is, as explained in \cite{BMbound} and reviewed in Appendix~B, to use a particular choice of field redefinition.
This choice  calibrates $s$ for a given theory to its Einstein value when compared at fixed values of $r_h$  or (equivalently) at fixed  temperature.
So that, in this field-redefinition frame, $s$ is set to its Einstein value irrespective of the of the theory, whether corrected or uncorrected by higher-derivative terms.

Using the proposed field redefinition, we find that
\be
\frac{\eta}{s} \;=\; \frac{\eta_E}{s_E}\left[1+\epsilon\tilde{b}(r_h)\right]
\;=\; \frac{1}{4\pi}\left[1+\epsilon\tilde{b}(r_h)\right],
\label{result}
\ee
where a subscript $E$ denotes the  Einstein value and
the tilde on $b$ indicates that it is the
previously defined coefficient but calculated in
the field-redefined  theory.

Let us note that this leading-order result is in agreement with
the  previous
methods (as discussed in Subsections~2.4-2.5), although our reasoning
is substantially different.

\section{Two-derivative effective actions of higher-derivative gravity}

The focus  of this section is on answering the following question:
What is the  two-derivative theory of gravity that effectively
describes, by way of  an action principle,  a given  higher-derivative model?
The answer involves
an important observation about the relationship between two-, three-
(and higher-) point functions and how this relationship can be used to further constrain the ratio $\eta/s$.

\subsection{Initial considerations}

To begin, let us recall
the effective two-derivative field equation for the $\{x,y\}$ gravitons,
\be
\left[1+\epsilon\tilde{b}(r)\right]\Box_E h_{xy} \;=\;0\;.
\label{BD2}
\ee
We wish to find a covariant action from which this equation is derived.
Before delving into the technical details, let us try to guess what the answer should look like.

As argued earlier, the effective theory should be a two-derivative covariant theory of small gravitational perturbations about a background solution. This is already enough to tell us that it must be one of the Lovelock theories. Then,  depending on the dimensionality of spacetime,
there are only a few possibilities. For $D=5$, the choices
are just Einstein and Gauss--Bonnet gravity. These considerations apply to the two- and three-point functions and, in a limited sense to be clarified later on,
also to the higher-point functions. Further simplification occurs when restricting attention to the
calculation of $\eta/s$ because, for any dimensionality,  Einstein and Gauss--Bonnet are then the only relevant theories. This was observed in \cite{BMratio} and
is  explained  in Appendix~C. Each of the two theories leads to a distinct value of $\eta/s$.

\subsection{A tale of two theories}

Let us first consider the  Einstein case.
We wish to find an ``Einstein-like'' Lagrangian whose
variation gives  Eq.~(\ref{BD2}). The answer is
\be
{\cal L}_{E}\;\propto\;
\left[1+\epsilon \tilde{b}(r)\right]\left({\cal R} +\frac{(D-1)(D-2)}{L^2}\right)
\;.
\label{Ein}
\ee
Here, $\tilde {b}(r)$  can be viewed as a radially dependent
correction to the gravitational coupling or an energy-scale dependent coupling from the perspective of the dual FT.
That is,
\be
\frac{1}{16\pi G_D}\;\to\;\frac{1}{16\pi G_D(r)}\;=\;
\frac{1}{16\pi G_D(r\to\infty)}\left[1+\epsilon \tilde{b}(r)\right]\;,
\ee
where $G_D$ is Newton's constant and we have used that higher-derivative corrections die off when $\;r\to\infty\;$.

The actual sense in which the coupling in (\ref{Ein}) is radially dependent is as follows: First,  the radius $r$ is fixed to some specific
value; for instance, the boundary of AdS or the horizon of the brane.
Then one calculates the 1PI amplitudes for that chosen radius. The amplitudes would then  be in agreement  to that of Einstein's theory
with the standard fixed Newton's constant.

In spite of appearances, the variation of Eq.~(\ref{Ein}) leads to
Eq.~(\ref{BD2}) without involving  derivatives of $\tilde{b}$.
Any resulting term in the variation of Eq.~(\ref{Ein}) contains two
or less derivatives, and so  a derivative on $\tilde{b}$ means
one or zero derivatives on the graviton. Such terms are
then (irrelevant) $\epsilon$-order mass terms.~\footnote{Any
term in the Lagrangian whose variation gives rise to a single derivative
on a graviton is, up to surface contributions,  equivalent to a mass term.}

Now, one  finds via the standard (two-derivative)  methods  that
$\eta$ and $s$ are both corrected
by the same factor of $1+\epsilon \tilde{b}(r_h)$, and so their ratio is (of course) the usual Einstein value
\be
\left[\frac{\eta}{s}\right]_{E} \;=\; \frac{1}{4\pi}\;.
\label{ratioE}
\ee

For the Gauss--Bonnet case, the calculation is technically more involved. As explained in Appendix~A, Eq.~(\ref{BD2}) would follow  from the variation of
\bea
{\cal L}_{GB}\; \propto\; {\cal R} &+&\frac{(D-1)(D-2)}{L^2}
\nonumber\\
&+&\epsilon L^2 \beta(r)
\left[{\cal R}^2 - 4{\cal R}_{ab}{\cal R}^{ab}
+{\cal R}_{abcd}{\cal R}^{abcd}\right]
\;,
\label{apx1}
\eea
where $\beta$ is a radial function such that, in the proximity
of the brane horizon,
\be
\beta(r) \;=\; -\frac {1}{2(D-1)(D-4)}\tilde{b}(r) \;.
\label{claim}
\ee
Again, $\beta$ can be viewed as a radially dependent coupling parameter
that should be fixed at a radius of choice.

One then finds that $\eta$ goes again  as
$1+\epsilon \tilde{b}(r_h)$.
However, now $s$ is uncorrected, as it would be for  any Lovelock theory
with an unmodified  Einstein--Hilbert term.
(This follows from the argument of Appendix~C, which is indifferent
to a  radially dependent coupling.) Hence,
\be
\left[\frac{\eta}{s}\right]_{GB} \;=\; \frac{1}{4\pi}
\left[1+\epsilon \tilde{b}(r_h)\right]\;.
\label{Gbound}
\ee
Depending on the sign of $\;\epsilon \tilde{b}(r_h)\;$,
the ratio $\;[\eta/s]_{GB}$ can be either smaller or larger than the Einstein value.

\subsection{The two, three (and higher) point functions}

Since   Einstein and  Gauss--Bonnet gravity
lead to different hydrodynamics, one may ask how else the two theories
are distinguished and whether such a distinction can be further used
to determine the ratio $\eta/s$.

Let us first suppose that there are no higher Lovelock terms, as would be the case for $D=5$. (We comment on Lovelock extensions at the end.)
Then,  once the value of $\eta/s$ has been determined to a given order in $\epsilon$, the graviton three-point function is determined to the very same order (as clarified below). In particular, the $n$-point functions of Einstein  gravity are polarization independent, as follows from the equivalence principle, whereas the $n$-point functions of a Gauss--Bonnet theory are polarization dependent in a specific way. Indeed,  this  is the very  same polarization dependence
that allows the value of $\eta/s$ to deviate from
the Einstein value   $1/4\pi$.

For the case of two-derivative theories, it was argued in
\cite{Mald-Hof,Hof} that  the two- and three-point functions are determined by a small number of interaction terms. Further, the two-point function is already enough to fix the three-point function, provided that the central charges of the FT dual are known \cite{fix}. The premise behind these  arguments can be understood as follows: The quota of at most two derivatives per term  and general covariance conspire to severely restrict how one can modify the propagator. If mass terms are disregarded (these can anyhow be absorbed in a redefinition of $\Box_E$), a modification of the linearized equations of motion is restricted to $\;\Box_E\to \Box_E +{\cal F}^{ab}\nabla_a \nabla_b \;$, where ${\cal F}^{ab}$ parametrizes the allowed corrections and its tensorial structure is determined by  the background geometry.

We  conclude that this intricate relation between the propagator and three-point
function can be used to constrain the choice of two-derivative theory.
A detailed presentation of  the Einstein and Gauss--Bonnet
 two- and three-point functions
(as well as their higher-point functions)
will be deferred to a subsequent publication, as the exact  expressions
 are not immediately useful.
Their basic forms, however,  can be deduced from  general covariance, the two-derivative limit on the linearized field equations, and the applicability of
field redefinitions and simplifying gauge conditions.

The end result is that the Einstein and Gauss--Bonnet
two-point functions can
respectively be expressed in terms of a few coefficients,
\be
\langle h_{ac} h_{bd} \rangle_E \;=\;
{\cal M}_{E}^{(0)} g^{ab}g^{cd}
h_{ac} h_{bd} \;+\;
{\cal N}_{E}^{(2)} g^{ab}g^{cd}g^{ef}
\nabla_e h_{ac} \nabla_f h_{bd}
\;,
\ee
\bea
\langle h_{ac} h_{bd} \rangle_{GB} &=&
\left({\cal M}_{GB}^{(2)}g^{ab}g^{cd}g^{ef}
+{\cal N}_{GB}^{(2)}{\cal R}^{cedf}g^{ab}
 +{\cal P}_{GB}^{(2)}{\cal R}^{acbd}g^{ef}
\right)
\nabla_e h_{ac}\nabla_f h_{bd}
\nonumber \\
&+&\left({\cal M}_{GB}^{(0)}g^{ab}g^{cd}
+{\cal Q}_{GB}^{(0)}{\cal R}^{acbd}
\right)
h_{ac}h_{bd}\;.
\eea
Here, the coefficients ${\cal M}$ and ${\cal Q}$ are functions of $r$ and the rest are numbers. The superscript 0 or 2 denotes the number of derivatives acting on the gravitons.

Similarly, the three-point functions can be expressed as
\be
\langle h_{ac}h_{be} h_{df}\rangle_E \;=\;
{\cal A}_{E}^{(0)}g^{ab}g^{cd}g^{ef}
h_{ac} h_{be} h_{df}\;+\;
{\cal B}_{E}^{(2)}g^{ab}g^{cd}g^{ef}g^{ij}
h_{ac}
\nabla_i
h_{be}
\nabla_j
h_{df}
\;,
\ee
\bea
\langle h_{ac}h_{be} h_{df}\rangle_{GB} &=&
\left({\cal A}_{GB}^{(2)}g^{ab}g^{cd}g^{ef}g^{ij}
+{\cal B}_{GB}^{(2)}{\cal R}^{eifj}g^{ab}g^{cd}+{\cal C}_{GB}^{(2)}{\cal R}^{cedf}g^{ab}g^{ij}
\right)
h_{ac}\nabla_i h_{be}\nabla_j h_{df}
\nonumber \\
&+&\left({\cal A}_{GB}^{(0)}g^{ab}g^{cd}g^{ef}
+{\cal D}_{GB}^{(0)}{\cal R}^{cedf}g^{ab}
\right)
h_{ac}h_{be} h_{df}\;,
\eea
where the coefficients ${\cal A}$ and ${\cal D}$ are functions of $r$ and the rest are numbers.

To simplify, we have
used that, for a theory of pure gravity, the
Ricci tensor  goes as \cite{spring}
$\;{\cal R}^{a}_{\ b}= \frac{1}{D}{\cal R}\delta^{a}_{\ b}\;$
and, for Einstein gravity only, the Riemann tensor reduces to
$\;{\cal R}^{ab}_{\;\;\;\;\;cd}= \frac{1}{D(D-1)}{\cal R}
\left[\delta^a_{\ c}\delta^b_{\ d}- \delta^a_{\ d}\delta^b_{\ c}\right]\;$.
Further,  we have
dropped any terms that are redundant via symmetries and assumed the
transverse--traceless
gauge  (extending to a general gauge would add more
terms but is otherwise straightforward).
It is also possible to express any $n$-point function for these particular theories using similar expressions, however, we will not do so in this paper.

The crucial point is that the Riemann tensor appears only in the
Gauss--Bonnet correlators.
The Einstein and Gauss--Bonnet $n$-point functions are, therefore,
distinguishable in way that cannot be eliminated by a field redefinition.

Now, given any higher-derivative theory, the simplest way to find the correct two-derivative description is to perform a field redefinition so as to bring the theory to the appropriate Lovelock one.
When the choice is strictly between  Einstein and Gauss--Bonnet,
it becomes the simple  matter of  inspecting the higher-derivative action or field equation for  polarization-dependent terms or, equivalently,
for the appearance of the four-index Riemann tensor ${\cal R}^{abcd}$.
If none appear,  the theory has a  natural completion to Einstein gravity and $\eta/s$ reverts to its Einstein value of $1/4\pi$.
Otherwise,  the theory  naturally completes  to Gauss--Bonnet gravity, leading to the value for $\eta/s$ as in Eq.~(\ref{Gbound}).

Both of these outcomes are compatible with our previous prescription (\ref{result}) in Section~2. This agreement is trivial when the choice is Gauss--Bonnet, so let us consider Einstein gravity which is a polarization-independent theory. Then, any field redefinition that changes $s$ must change all two-point couplings by the same amount. Hence, for the Einstein case, $s\to s_E$ implies that $\eta\to\eta_E$ or $b\to\tilde{b}=0$.

For the purpose of evaluating higher-point functions, additional input is necessary. This is on account of possible Lovelock extensions when $D>6$ and, irrespective of $D$, information about the higher-point functions that is lost in the linearized theory.
So that, in general, further information besides polarization dependence versus independence is needed to clarify the precise identity
of a polarization-dependent theory. Such additional input could be a perturbative hierarchy between the coefficients of the interaction terms. For example,
suppose that  the terms quadratic in Riemann tensor are of order $\epsilon$ while those quartic in the Riemann tensor are of order $\epsilon^2$. Then the magnitude of the deviation of the ratio $\eta/s$ from its Einstein value  $1/4\pi$
and, similarly, the magnitude of the deviation of the three-point function from its Einstein value can be used to find out which term is responsible for the leading correction. Once this leading contributor is identified, then its
contribution to four- and higher-point functions in some kinematic region can also be determined to leading order in $\epsilon$.

\section{Conclusion}

To summarize, we have proposed that gravitational perturbations about a background solution of a unitary and UV-complete theory  should be described by an effective theory of a single massless graviton whose linearized equations of motion contain at most  two derivatives. Given some higher-derivative corrections to Einstein gravity, we have explained how to find this theory by imposing a novel set of boundary conditions that ensures unitarity and removes any unwanted dependence on the cutoff-scale physics.  In particular, our procedure assigns  an unambiguous value to $\eta/s$ for any  higher-derivative model.

For the purpose of calculating $\eta/s$,
there are only two possible candidates for a two-derivative description
of higher-derivative theories; either Einstein or Gauss--Bonnet gravity.
We have provided an explicit criterion for choosing among the pair that is based on the polarization dependence of their 1PI on-shell amplitudes. It has been verified that the resulting value of $\eta/s$ agrees with our earlier prescription and with previous calculations in the literature. We have stressed the issues with directly calculating
quantities like $\eta$ and $s$ in a truly higher-derivative framework.

To determine the 4- and higher-point functions, in general, further information besides polarization dependence versus independence is needed.  However, if one is satisfied
with their leading-order values, then the leading-order correction to $\eta/s$ is sufficient to determine all higher-point functions to leading order, as explained in the last paragraph of Section~3.

We anticipate that our proposals can be compared with properties of strongly coupled fluids such as the quark-gluon plasma or cold atoms.
As discussed in \cite{Mald-Hof,Hof}, the LHC should be able to measure
energy correlation functions for  a strongly
coupled fluid. If this fluid has a gravitational dual, these correlators
amount to graviton scattering amplitudes.
Then the polarization sensitivity of the Gauss--Bonnet three-point function should be manifest through an angular dependence that is absent in the Einstein case.

\section*{Acknowledgments}

We are indebted to Ofer Aharony for his help and advice, for many valuable and detailed discussions and for many useful suggestions. We wish to thank Elias Kiritsis for valuable suggestions and comments on the manuscript and to thank Cliff Burgess and Shanta de Alwis for useful discussions.
The research of RB was supported by the Israel Science Foundation grant no. 239/10. The research of AJMM received support from the Korea
Institute for Advanced Study (KIAS) and Rhodes University. AJMM thanks the Department of Physics, Ben-Gurion University and the Theory Division, CERN for their hospitality.

\appendix

\section{The Gauss-Bonnet Lagrangian}
\label{AA}

In this appendix, we verify  Eqs.~(\ref{apx1}) and (\ref{claim}) in the main text.

Let us begin by considering just the four-derivative part  of
the (generalized) Gauss--Bonnet Lagrangian in Eq.~(\ref{apx1}),
\be
\beta {\cal L}_{\Delta GB} \;\equiv \; \beta L^2
\left[{\cal R}^2 - 4{\cal R}_{ab}{\cal R}^{ab}
+{\cal R}_{abcd}{\cal R}^{abcd}\right]\;,
\ee
where $\beta=\beta(r)$ is unspecified and we have dropped
 $\epsilon$.

Inserting this Lagrangian into the $x,y$ component of
the generic field equation (\ref{wald-EQ}),
we have
\be
\beta{\cal R}_{abcx}\left[\frac{\delta{\cal L}_{\Delta GB}}
{\delta{\cal R}_{abc}^{\;\;\;\;\;y}}\right]
-2\left[\frac{\delta{\cal L}_{\Delta GB}}
{\delta{\cal R}_{a\;\;\;\;b}^{\;\;xy\;\;}}\right]
\nabla_a\nabla_b \beta
\;=\; \frac{1}{2}g_{xy}\beta{\cal L}_{\Delta GB}\;.
\label{apx2}
\ee
This form follows from a special property of any Lagrangian
${\cal L}_{LL}$ of the Lovelock class
(for a formal proof, see Appendix 3 in \cite{Pad}):
\be
\nabla_a \left[\frac{\delta{\cal L}_{LL}}
{\delta{\cal R}_{abcd}}\right]
\;=\; 0\;.
\ee

The next step is to linearize Eq.~(\ref{apx2}) and  then extract the  two-derivative terms. The linearization of the right-hand side produces a term proportional to $h_{xy}$,  so that this is just a mass term.
What remains simplifies greatly in the transverse-traceless gauge,
as is appropriate for
the $h_{xy}$ gravitons once the radial gauge has been imposed.
The left-most term can be evaluated in a tedious but straightforward manner.
It can then be expressed as
\be
\beta {\cal G}\Box h_{xy} \;,
\ee
where we have defined
\be
{\cal G} \;\equiv \; 4 L^2 \left[\frac{D-2}{2D}{\cal R}
-{\cal R}^x_{\ x}  -(D-4){\cal R}^{xy}_{\;\;\;\;\;xy}
\right]\;.
\label{calG}
\ee

For sake of completeness, the above expression follows from
the kinetic part of the linearized  field equation or
$\; {\cal G}^{ab} \Box h_{ab}\;$,
for which
\be
{\cal G}^a_{\ b}\;\equiv\;
 4L^2\left[ \delta^a_b \left(\frac{1}{2}{\cal R}
-\sum^{\{x,y\}}_{c}{\cal R}^c_{\ c} + \frac{1}{2}
\sum_{c\neq d}^{\{x,y\}}{\cal R}^{cd}_{\;\;\;\;\;cd}\right)
+ \sum_{c}^{\{x,y\}}{\cal R}^{ac}_{\;\;\;\;\;\;bc}
-{\cal R}^{a}_{\;\;b}\right]
\;.
\label{thingsx}
\ee
To obtain Eq.~(\ref{calG}), we have set  $\;2{\cal G}={\cal G}^{xy} +{\cal G}^{yx}\;$
and then used symmetries of the background to simplify the form of ${\cal G}$.

Meanwhile, the second term in Eq.(\ref{apx2}) works out to be
\be
2\epsilon L^2\left[g^{ab}\Box h_{xy}-\nabla^a\nabla^b h_{xy}\right]\nabla_a\nabla_b \beta\;.
\ee

We now want to ``reverse engineer'' so as  to produce the desired result in
Eq.~(\ref{BD2}).
So let us fix
\be
{\cal G}\beta\Box h_{xy}
+2L^2\left[g^{ab}\Box h_{xy}-\nabla^a\nabla^b h_{xy}\right]\nabla_a\nabla_b \beta
\;=\;
\tilde{b}\Box_E h_{xy} \;.
\label{what}
\ee

To solve the previous equation for the unknown function $\;\beta(r)\;$, one can impose the
hydrodynamic limit, whereby only the radial differentiations of $h_{xy}$ turn out
to be relevant.
This is enough to eliminate the $\nabla\nabla\beta$ term.
We also take the near-horizon limit, so that $\;\Box_E\to\Box\;$.
This leads to
\be
{\cal G}\beta\Box h_{xy}
\;=\; \tilde{b}\Box h_{xy} \;
\label{what2}
\ee
or
\be
\beta \;=\; {\cal G}^{-1}\tilde{b}\;.
\ee

Finally, inserting  the horizon value of $-2(D-1)(D-4)$
for ${\cal G}$, we arrive at
\be
\beta \;=\; -\frac{1}{2(D-1)(D-4)}\tilde{b}\;,
\ee
as claimed.

\section{Review of the scaling properties of $\eta$ and $s$ }
\label{AC}

Here, we review an argument
from \cite{BMbound} that is referred to in the main
text.

Let us consider some generic theory of gravity $X$ in a $D$-dimensional
AdS spacetime
with a black brane solution.
Our interest is in the  entropy density $s_X$,
which can  be obtained by way of Wald's formalism \cite{wald1,wald2}.
Then, via the analysis of \cite{BGH-0712.3206}, $s_X$ can
be expressed in the form of the area law,
\be
s_{X}\;=\;\frac{\sigma}{4G_{X}}\;,
\ee
where $\;\sigma\sim r_h^{D-2}$ is the area density of the
brane~\footnote{We have left $\sigma$ unlabeled, as it is assumed
to be fixed for the purpose of comparing theories.
This is equivalent to comparing theories at fixed $r_h$ or fixed temperature.}
and $G_{X}$ is related to but generally different
than the $D$-dimensional Newton's constant $G_E$
(the subscript $E$ labels quantities in Einstein's theory).

Now, as shown  in \cite{BMbound}, there exists a
constant conformal transformation of the metric (along with an accompanying
rescaling of  $G_{X}$ that  is necessary to preserve
the form of the Einstein term in the action) such that
\be
s_{\widetilde X}\;=\;\frac{\sigma}{4G_E}\;,
\ee
with ${\widetilde X}$ denoting the transformed theory. That is, the
conformal transformation is chosen specifically to ensure
\be
G_{X}\;\to\; G_{\widetilde X}\;=\; G_{E}\;.
\ee

That the entropy density can be freely changed
may seem odd, but this is because densities such as
$s$ are ambiguously defined.
Conversely, the entropy $\;S=sV_{\perp}\;$  does not change under such a
transformation ($V_{\perp}$ is the transverse volume of the brane).
Similarly, the ratio of two densities cannot change, and so
\be
\frac{\eta_{\widetilde X}}{s_{\widetilde X}}\;=\; \frac{\eta_X}{s_X}\;.
\ee

The advantage of using the transformed system is that now
 $\;s_{\widetilde X}=s_E\;$ (in units of fixed temperature), and
so the net correction to the ratio $\eta_{\widetilde X}/s_{\widetilde X}$
is entirely captured by the viscosity.

\section{Review of the Lovelock calculation}

Finally, we recall
from \cite{BMratio} why
higher-order Lovelock terms cannot contribute
to the ratio $\eta/s$.

Let us first consider the entropy density. It is already known that the $m^{{\rm th}}$-order
term ${\cal L}_m$  in the Lovelock Lagrangian
(where ${\cal L}_0 $ is a constant,
${\cal L}_1$ is Einstein, ${\cal L}_2$ is Gauss--Bonnet, {\it etc.}),
makes a contribution
to the entropy or its density that, up to constant factors, goes as
${\cal L}_{m-1}[g_{\parallel}]$ \cite{JM,MV}.
Here, $g_{\parallel}$ has a special meaning:
One should incorporate only curvature components of the form
${\cal R}^{x_ix_j}_{\;\;\;\;\;x_ix_j}$ (where $x_i$ and $x_j$ are
any orthogonal pair of transverse brane directions)
and permutations thereof. Notably,
any such curvature component vanishes at the horizon of a black brane;
so that the only non-vanishing contribution to $s$ is from the
$m=1$ Einstein term,
for which $\;{\cal L}_{m-1}[g_{\parallel}]={\cal L}_0\;$ is a constant.

By direct analogy, the contribution to $\eta$ from  the order-$m$ term
goes as ${\cal L}_{m-1}[g_{\perp}]$ \cite{BMratio}, where  $g_{\perp}$
indicates  that one should now only  incorporate the curvature
component ${\cal R}^{rt}_{\;\;\;\;\;rt}$. But, since the background geometry
only varies with $r$, one is effectively calculating the $m-1$ Lovelock
term for a two-dimensional theory. As is well known,  the $p^{\rm th}$-order
term in the Lovelock Lagrangian vanishes identically when $\;p>D/2\;$.
Then, since the dimension is effectively $D=2$,
  we have no contribution to $\eta$ when $\;m-1>2/2\;$ or
$\;m>2\;$. Meaning that, for the shear viscosity, only the Einstein and
Gauss--Bonnet terms are contributors. (See, {\it e.g.},
\cite{verify} for an explicit verification.)

\end{document}